# Numerical simulation of individual coil placement – A proof–of–concept study for the prediction of recurrence after aneurysm coiling

Julian Schwarting[1]; Fabian Holzberger[2]; Markus Muhr[2], Martin Renz[1]; Tobias Boeckh-Behrens[1]; Barbara Wohlmuth[2], Jan Kirschke[1]

1    Department of Diagnostic and Interventional Neuroradiology, Technical University of Munich, Germany, School of Medicine, Klinikum rechts der Isar, Germany
2    Technical University of Munich, School of Computation, Information and Technology, Department Mathematics.

**Corresponding author:**
Dr. Julian Schwarting, MD, B.Sc.
Technical University of Munich, Germany, School of Medicine, Klinikum rechts der Isar, Ismaninger Straße 22, 81675 München Germany
E-Mail: Julian.Schwarting@tum.de
Phone: +49 89 4140 4652

**Abstract**

Rupture of intracranial aneurysms results in severe subarachnoidal hemorrhage, which is associated with high morbidity and mortality. Neurointerventional occlusion of the aneurysm through coiling has evolved to a therapeutical standard. The choice of the specific coil has an important influence on secondary regrowth requiring retreatment. Aneurysm occlusion was simulated either through virtual implantation of a preshaped 3D coil or with a porous media approach. In this study, we used a recently developed numerical approach to simulate aneurysm shapes in specific challenging aneurysm anatomies and correlated these with aneurysm recurrence 6 months after treatment. The simulation showed a great variety of coil shapes depending on the variability in possible microcatheter positions. Aneurysms with a later recurrence showed a tendency for more successful coiling attempts. Results revealed further trends suggesting lower simulated packing densities in aneurysms with reoccurrence. Simulated packing densities did not correlate with those calculated by conventional software, indicating the potential for our approach to offer additional predictive value. Our study, therefore, pioneers a comprehensive numerical model for simulating aneurysm coiling, providing insights into individualized treatment strategies and outcome prediction. Future directions involve expanding the model's capabilities to simulate intraprocedural outcomes and long-term predictions, aiming to refine occlusion quality criteria and validate prediction parameters in larger patient cohorts. This simulation framework holds promise for enhancing clinical decision-making and optimizing patient outcomes in endovascular aneurysm treatment.

**Introduction**

Aneurysmatic subarachnoid hemorrhage (aSAH) is a severe type of hemorrhagic stroke that results from a rupture of intracranial aneurysms.[1] Despite a global decline in the incidence of aSAH attributed to public health initiatives and lifestyle modifications, the associated mortality rates persist at elevated levels.[2] Brain aneurysms are focal dilations of major intracranial arteries arising from the structural weakening of the vessel wall. Aneurysms have a prevalence of approximately 3% in a population without comorbidity, which varies strongly between geographical regions, age, and sex.[3,4]

Interventional occlusion of ruptured or incidentally detected brain aneurysms has evolved significantly over the last decades and is even associated with better disability-free survival 1 year after aneurysm rupture compared to neurosurgical clipping. However, depending on the individual case, both are equally feasible therapeutic options.[5]

Within the last decades, several neurointerventional methods were developed for aneurysm occlusion. In general, intraaneurysmatic occlusion strategies (i.e., Coiling, Web Devices, and Contour devices) can be distinguished from extraaneurysmatic occlusion strategies (i.e., Flow diverter, Stenting).[6] Endovascular coiling is still the most used technique for interventional aneurysm occlusion, as it is widely available, has comparably low costs, can be used in various aneurysm shapes, and provides immediate protection from re-rupture after SAH. Here, the aneurysm cavity is filled with different platinum coils, triggering an intraaneurysmatic thrombus formation within a few minutes.[7] After the intervention, the endothelium re-grows to complete a successful aneurysm occlusion within months. A secondary consolidation of the coil package through compactification of the thrombus, inflammatory processes, and/or high stresses in the aneurysm walls can, however, result in a regrowth of the aneurysm. Therefore, the choice of the material highly influences the overall treatment success. Coils vary significantly in length, diameter, shape, and material and are commercially available from several manufacturers.[7] Despite specialized endovascular procedures and advanced imaging techniques, device selection is mostly governed by the personal experience of the neuroradiologist.[7-9] However, due to the great variety of available coils, the optimal choice of the coil may be difficult in challenging cases, and an objective simulation of the coiling procedure with outcome prediction may result in lower rates of re-occurrence.

We, therefore, simulated the specific shape of the actual coils used for coiling of 6 aneurysms in a proof–of–concept study in patients with and without aneurysm reoccurrence and investigated whether the probability of correct coil insertion or the packing density was associated with the occurrence of aneurysm regrowth.

## Methods

### Study design and patient selection

We retrospectively examined the medical data of all patients who were endovascularly treated at our center for brain aneurysms between January 2013 and December 2016 with a single coil and who had a regrowth of the aneurysm that was observed in a follow–up angiography after 6 months. For comparative analysis, we selected 3 cases of anterior circulation aneurysms based on size and shape, which did not show aneurysm reoccurrence in the follow-up angiography.

### Clinical data

Clinical data was retrospectively extracted from clinical files of selected patients. Data is fully anonymized. The regrowth of aneurysms was classified by the Raymond & Roy classification after approximately 6 months.[10] Aneurysms were characterized by size and location. Locations were categorized into middle cerebral artery (MCA), anterior cerebral artery (ACA), anterior communicating artery (ACOM), and internal carotid artery (ICA). The packing density of the aneurysm after the coiling procedure was calculated using the commercially available software AngioSuite (Cascade Medical, Knoxville, TN, USA)[11], which calculates aneurysm volumes from 2D images.

**Radiological data selection and workflow of simulation**

Virtual coiling was performed in images extracted from a 3D rotational angiography (Philips Azurion 7 Neuro Suite; Philips Medical Systems, Netherlands; 3.7 pixels / mm) acquired for the coiling procedure. Images were semiautomatically to a 3D model using ITK SNAP with a combination of a threshold-based and a region-growing algorithm [12]. An experienced neuroradiologist validated the geometry. Virtual coiling was performed using our recently published numerical model.[13]

In summary, coiling was individualized based on the specific coils used in the investigated case. We simulated the coil structure for every coil, identified by its helical shape, characteristic length, and the diameters D1 - D3: *D1* of the coil stock wire; *D2*, the radius of the helical shape; *D3* representing the macroscopic shape. (**Fig.1**)

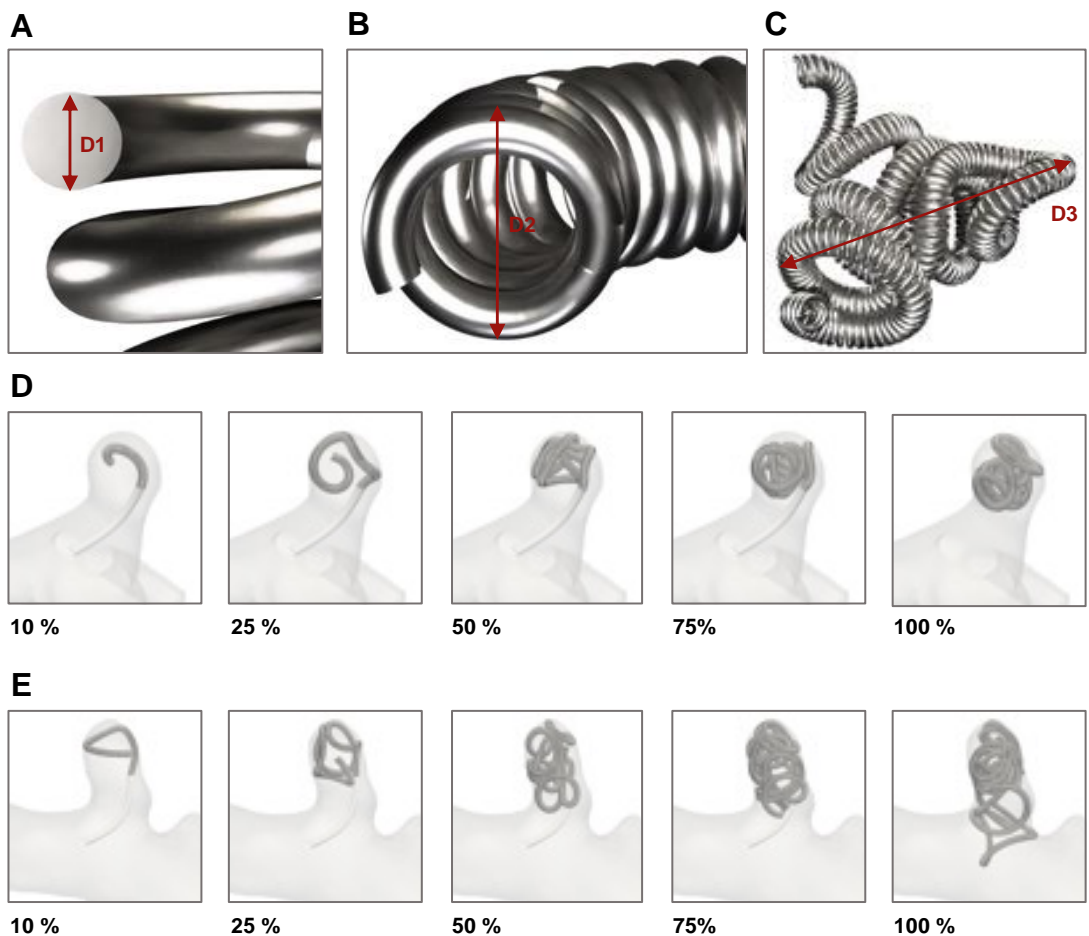

*Figure 1* – *Inserted coils were characterized using Diameters D1 of the coil stock wire (A) D2, the radius of the helical shape(B) and D3 representing the macroscopic shape (C). Coiling attempts were either classified as successful (A) or unsuccessful (B) based on the Raymond & Roy classification and potential protrusion into the carrier vessel.*

The properties of each coil were taken from the publicly available datasheet. During the virtual coiling, the microcatheter was placed in the aneurysm from which the coil was extended. An experienced neurointerventionalist set the optimal realistic position of the microcatheter. To evaluate the technical difficulty of placing the coil correctly, we simulated 100 coiling procedures per aneurysm. Here, the microcatheter tip was randomly repositioned within a sphere with a radius of 0.5 mm around the optimal position.

**Evaluation of virtual coiling simulation**

The success of the virtual coiling procedure was evaluated in 3 outcome parameters:

1. Packing density of the aneurysm, which was calculated as a percentage of the aneurysm volume

2. Volume of the coil protruding beyond the neck into the carrier vessel in % of the total coil volume

3. Clinical evaluation of the coiling result, assessed in a 3D model by an experienced neuroradiologist. A successful virtual coiling attempt is shown in **Fig. 1D,** and an unsuccessful virtual coiling is shown in **Fig. 1E**

**Ethical approval**

The study was approved by the local Research Ethics Committee of the Technical University of Munich (186/20S). Informed consent was not applicable due to the retrospective design in accordance with the standards of the local Ethics Committee. The study was performed in accordance with the guidelines of the Declaration of Helsinki.

**Statistical analyses**

Due to small sample numbers, we did not assume normal distribution of the data. Descriptive statistics such as the median and interquartile range (IQR) are provided. Statistically significant differences between groups were tested with the Mann-Whitney test for non-parametric data using Prism 10 (Graphpad Software LLC, USA) with a two-sided significance level (p) of 0.05.

# Results

## Patient characteristics

6 patients were included in the study. 3 patients with an ACOM-Aneurysm, 1 case of an ICA aneurysm, and 1 case of an MCA aneurysm with diameters of 1.5 – 6.5 mm. 2/3 of the included aneurysms were ruptured. Postprocedural packing density calculated by procedure times varied between 38 and 155 minutes. In follow-up, 3 aneurysms showed no signs of reoccurrence (Raymond & Roy 1), 1 patient had an aneurysm reoccurrence classified as Raymond & Roy 2, and 2 patients had an aneurysm reoccurrence classified as Raymond & Roy 3B. **(Tab. 1)** *Case 3* was coiled stent-assisted and had a Coil volume that was protruding in the lumen of the carrier vessel.

| Table 1 – Baseline characteristics | | | | | | | | |
|---|---|---|---|---|---|---|---|---|
| Case | Location | Rupture | Aneurysm diameter (mm) | Inserted Coil | Procedure duration (min) | Packing density (%) | Initial occlusion (R&R) | 6 month Follow-up (R&R) |
| 1 | ACOM | Yes | 1.5 | *Stryker Target Helical Ultra, D3 = 1 mm, L = 2 cm* | 38 | 8 | R&R I | R&R I |
| 2 | ACOM | Yes | 6.5 | *Stryker Target 360 Ultra* | 63 | 31 | R&R I | R&R I |
| 3 | MCA | No | 1.7 | *Stryker Target Helical Nano, D3 = 1 mm, L = 2 cm* | 155 | 168 | R&R I | R&R I |
| 4 | ACA | Yes | 3.8 | *Stryker Target Helical Ultra, D3 = 2.5 mm, L = 6 cm* | 138 | 33 | R&R I | R&R II |
| 5 | ICA | No | 4.7 | *Stryker Target 360 soft, D3 = 4 mm, L = 8 cm* | 116 | 19 | R&R I | R&R III B |
| 6 | ACOM | Yes | 2.5 | *Stryker Target Helical Ultra, D3 = 1.5 mm, L = 3 cm* | 42 | 58 | R&R III B | R&R III B |

*Note – ACOM = Anterior communicating artery; MCA = Middle cerebral artery; ICA = Internal carotid artery; R&R = Raymond & Roy classification*

## Virtual coiling results

Coiling results with the direct comparison of the real angiographic result immediately after coiling and in the first follow-up angiography in a case without recurrence (Raymond & Roy class I, *Case 1*), with a residual aneurysm neck (Raymond & Roy class II, *Case 4*) and the occurrence of a new contrast opacification outside the coil mesh (Raymond & Roy class IIIb; *Case 5*) are shown in **Fig. 2.**

**Case 2**

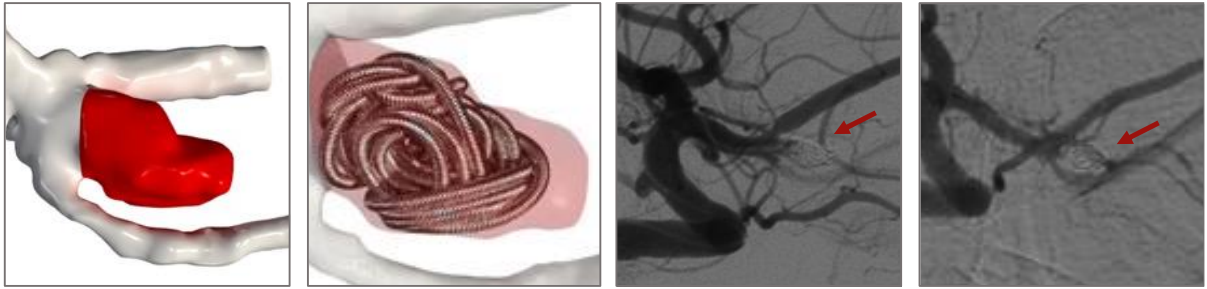

**Case 4**

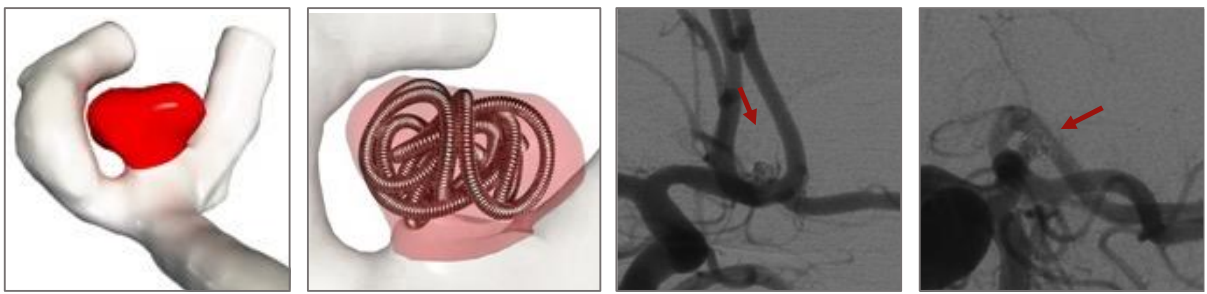

**Case 5**

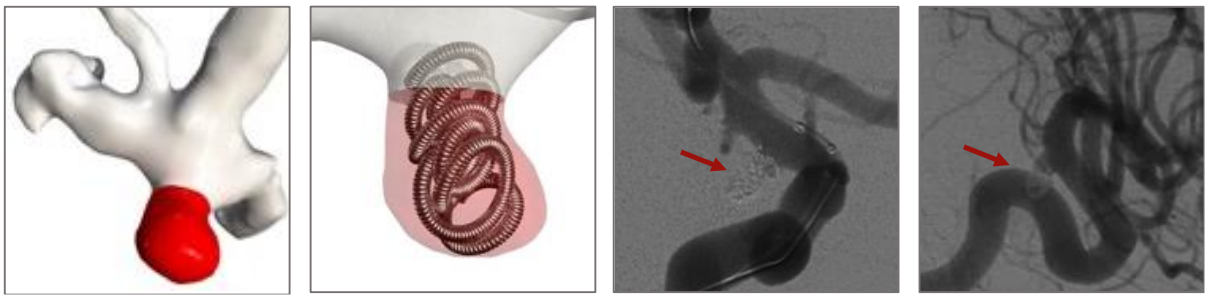

*Figure 2 – Simulation of a coiling in a specific aneurysm. Angiographic screenshots show results immediately after coiling (left) and 6 months after coiling (right). Case 2 had a full occlusion after 6 months (R & R 1), Case 4 had a coil compactification with a neck remnant after 6 months (R & R 2) and Case 5 had an aneurysm reoccurrence classified (R & R 3B)*

Simulated coiling was successful in 16 – 100% of all attempts with the coil used in the real scenario. Simulated packing density varied between 15.82 and 24.02%. **(Tab. 2)**

Simulated packing density showed a non-significant trend of lower densities in aneurysms without secondary reoccurrence (**Fig. 3A**) and did not correlate to the angiographic packing density calculated by AngioSuite (**Fig. 3B**).

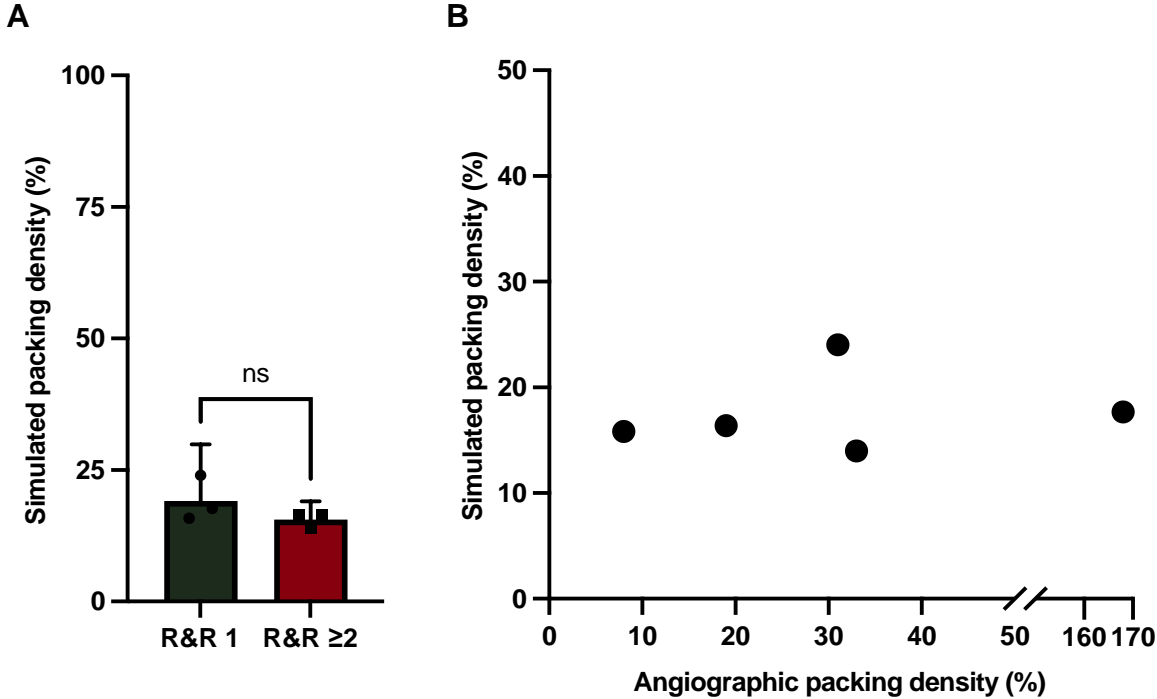

*Figure 3* –

*(A) Simulated packing density in aneurysms without and with reoccurrence*

*(B) Correlation of simulated packing densities and packing densities calculated by AngioSuite*

| Table 2 – Simulation results | | | |
|---|---|---|---|
| Case | Simulated Packing density Median [IQR] in % | Successful Coiling attempts (%) | Simulated Extraaneurysmal Volume Median [IQR] in % |
| 1 | 15.82 [0.84] | 69 | 4.27 [5.12] |
| 2 | 24.02 [0.34] | 64 | 0.44 [1.37] |
| 3 | 17.68 [0.01] | 100 | 0 [0] |
| 4 | 13.98 [0.56] | 78 | 3.88 [4.10] |
| 5 | 16.39 [1.24] | 16 | 9.24 [6.95] |
| 6 | 16.41 [0.48] | 29 | 0.04 [2.99] |

The proportion of successful coiling attempts as a surrogate parameter for variability in possible microcatheter positions allowing for successful coiling was lower in aneurysms with secondary reoccurrence (**Fig. 4A**), and the extraaneurysmatic coil volume was larger in these cases (**Fig. 4B**); these results were, however, not statistically significant.

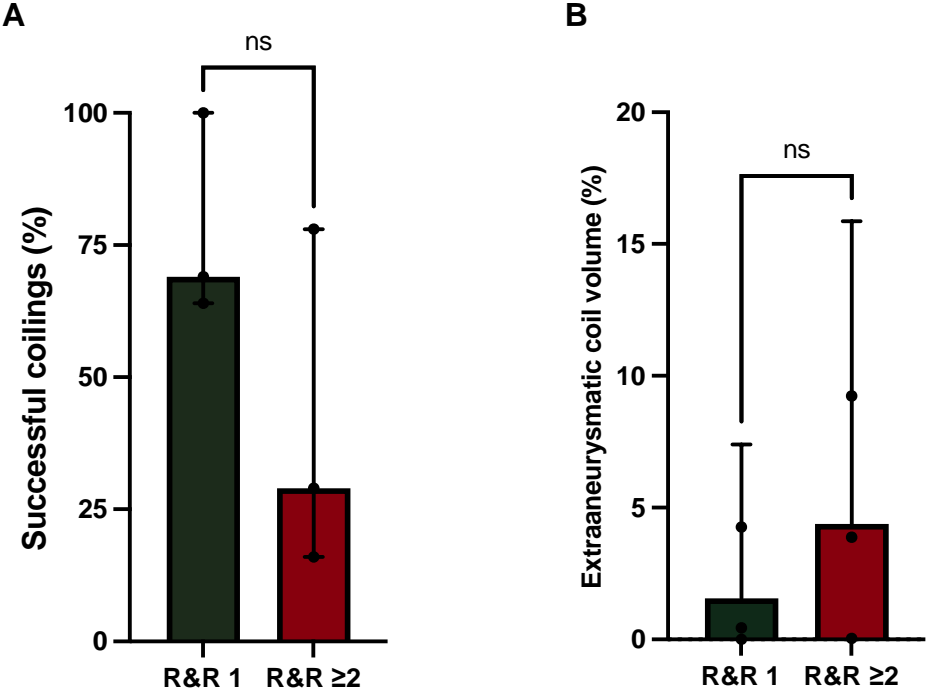

*Figure 4*

*(A) Successful coiling attempts* as a surrogate parameter for variability in possible microcatheter positions allowing for successful coiling *in aneurysms without or with reoccurrence*
*(B) Extraaneurysmatic coil volume in aneurysms without and with reoccurrence*

**Discussion**

In this proof-of-concept study, we implemented a workflow of virtual coiling in 6 challenging real cases of intracranial aneurysms with or without reoccurrence in the follow-up angiography.
Our numerical model simulates the specific shape of the implanted coil used during the intervention in the actual geometry of the investigated case. We observed trends that show lower clinically sufficient coiling attempts in aneurysms with lower simulated packing densities so that this parameter could play a major role in outcome prediction after validation in larger datasets.
The investigation of the simulation of occlusion techniques for intracranial aneurysms has increased in recent years.[12 14-17] Due to the limited resolution of available imaging methods, simulations were predominantly based on porous media models, which are limited in their clinical use as they do not provide a clear recommendation for the use of a specific neurointerventional device for the clinician.[18]

To our knowledge, there is only one software package available for clinical use, which simulates a preshaped 3D model of a neurointerventional device in the aneurysm sac or the carrier vessel. [17]

With our numerical approach, [13] we provide a novel tool that can predict the exact shape of the implanted coil in a patient-specific anatomy. This approach could provide additional information for neurointerventionalists in the selection of the best coil and can also be used as part of a complete in-silico simulation of an aneurysm coiling to reduce complications and retreatment in the long term.

Through a varying placement of the microcatheter tip, we could observe strong variations in the shapes of the implanted coils, a phenomenon commonly known in clinical practice that has not been considered in other coil simulations. Interestingly, we observed a trend that the variability in possible microcatheter positions allowing for a successful coiling might be associated with an aneurysm recurrence and, therefore, could potentially be used for outcome prediction.

Simulated packing densities did not correlate with the calculated packing densities from the AngioSuite software package[11] which is already available for clinical use. Our simulated densities were lower than those of the automated software package. This could be attributed to the differing aneurysm volumes calculated by AngioSuite from 2D images or the high resolution of our simulated coil: Aneurysm volume strongly depends on the volume reconstruction and the image resolution and was validated by an experienced neuroradiologist in our dataset. The coil volume was simulated to the base wire in our model. However, the resolution used for the coil volume calculation is not reported in AngioSuite. Additionally, we only considered intraaneurysmatic coil volume for our measurements. Interestingly, we could also not observe correlations between packing densities calculated in AngioSuite and aneurysm reoccurrence in a previous study.[19]

The numerical model shown in this study is the foundation for a comprehensive model for aneurysm coiling simulation. The simulation will be tested in larger cohorts and will be extended to be even more applicable for clinical use. In the future, further adaptations of the model will be implemented to increase the accuracy of the simulation further:

**Intraprocedural outcome simulation**

In the current model, the vessel wall is rigid. An elastic wall and wall pulsation could take a potential deformation of the aneurysm or the carrier vessel into account, which can potentially influence the coil shape. The simulation of multiple coils would extend the applicability of the simulation, especially for large aneurysms.

**Longterm outcome prediction**

For long-term outcome prediction, intraaneurysmatic thrombus formation likely would likely result in a more realistic simulation of the intraaneurysmatic packing density. The implication of a flow simulation with pulsatile flow velocities and the investigation of vessel wall inflammation

**Conclusion**

The numerical simulation of aneurysm coiling shown in this study provides an individual coiling simulation to provide additional information for coil selection. These first promising results show the possible benefit of a coiling simulation framework in clinical applications. The next steps include a more general coiling model, e.g., through multiple coils and elastic vessel walls. A larger patient cohort is required to validate the approach and develop refined occlusion quality criteria.

**Availability of data and materials**

All source data are stored at the Department of Diagnostic and Interventional Neuroradiology, Technical University, Munich. We invite parties interested in collaboration and data exchange to contact the corresponding author directly.

**Funding**

Fabian Holzberger, Markus Muhr and Barbara Wohlmuth gratefully acknowledge the financial support provided by the Deutsche Forschungsgemeinschaft under the grant number WO 671/11-1. Julian Schwarting, Fabian Holzberger, Markus Muhr, Barbara Wohlmuth and Jan Kirschke gratefully acknowledge the financial support provided by the Deutsche Forschungsgemeinschaft project number 465242983 within the priority programme "SPP 2311: Robust coupling of continuum-biomechanical in silico models to establish active biological system models for later use in clinical applications - Co-design of modeling, numerics and usability" (WO 671/20-1)